\documentclass[%
reprint,
superscriptaddress,
showkeys,
nofootinbib,
amsmath,amssymb,
aps,
pra,
]{revtex4-2}

\usepackage{graphicx}
\usepackage{dcolumn}
\usepackage{bm}
\usepackage{hyperref}
\usepackage{verbatim}
\usepackage[utf8]{inputenc}
\usepackage{layouts}
\usepackage{multirow}
\usepackage[dvipsnames]{xcolor}
\usepackage{hhline} 
\usepackage[separate-uncertainty=true]{siunitx}
\usepackage[version=3]{mhchem} 
\usepackage{wrapfig}
\usepackage{braket}
\usepackage{float}
\usepackage{leftidx} 
\usepackage{amsmath,mathtools}
\usepackage{amssymb}
\usepackage{chngcntr}
\renewcommand{\figurename}{\textbf{Figure}}

\newcommand{\moire}{moir\'e }
\newcommand{\cd}[1]{c^{\dagger}_{#1}}
\newcommand{\co}[1]{c^{}_{#1}}

\newcommand{\fd}[1]{f^{\dagger}_{#1}}
\newcommand{\fo}[1]{f^{}_{#1}}

\hypersetup{colorlinks=true,urlcolor=magenta,linkcolor=magenta,citecolor=cyan}

\newcommand{\TUM}{\affiliation{Technical University of Munich, TUM School of Natural Sciences, Physics Department, 85748 Garching, Germany}}

\newcommand{\CIT}{\affiliation{Technical University of Munich, TUM School of Computation, Information and Technology, 80333 Munich, Germany}}

\newcommand{\WSI}{\affiliation{Walter Schottky Institute, Technical University of Munich, 85748 Garching, Germany}}

\newcommand{\MCQST}{\affiliation{Munich Center for Quantum Science and Technology (MCQST), Schellingstr. 4, 80799 M{\"u}nchen, Germany}}

\renewcommand{\thefootnote}{\fnsymbol{footnote}}
\newcounter{emailcounter}
\stepcounter{emailcounter}

\begin{document}
\title{Gate-tunable Bose-Fermi mixture in a strongly correlated moiré bilayer electron system}

\author{Amine Ben Mhenni$^{\fnsymbol{emailcounter},\stepcounter{emailcounter} \fnsymbol{emailcounter}}$}

\WSI
\TUM
\MCQST

\author{Wilhelm Kadow$^\fnsymbol{emailcounter}$}%

\TUM
\MCQST

\author{Mikołaj J. Metelski}%
\WSI
\TUM
\MCQST

\author{Adrian O. Paulus}%
\WSI
\TUM
\MCQST

\author{Alain Dijkstra}%
\WSI
\TUM
\MCQST

\author{Kenji Watanabe}%
\affiliation{%
 Research Center for Electronic and Optical Materials, National Institute for Materials Science, 1-1 Namiki, Tsukuba 305-0044, Japan
}%

\author{Takashi Taniguchi}%
\affiliation{%
 Research Center for Materials Nanoarchitectonics, National Institute for Materials Science,  1-1 Namiki, Tsukuba 305-0044, Japan
}%

\author{Seth Ariel Tongay}%
\affiliation{%
 Materials Science and Engineering, School of Engineering for Matter, Transport, and Energy, Arizona State University, Tempe, AZ USA 85287
}%

\author{Matteo Barbone}%
\WSI
\CIT
\MCQST

\author{Jonathan J. Finley$^{\stepcounter{emailcounter}\fnsymbol{emailcounter}}$}%
\WSI
\TUM
\MCQST

\author{Michael Knap$^{\stepcounter{emailcounter}\fnsymbol{emailcounter}}$}%
\TUM
\MCQST

\author{Nathan P. Wilson$^{\stepcounter{emailcounter}\fnsymbol{emailcounter}}$}%
\WSI
\TUM
\MCQST

\date{\today}

\begin{abstract}

Quantum gases consisting of species with distinct quantum statistics, such as Bose-Fermi mixtures, can behave in a fundamentally different way than their unmixed constituents. This makes them an essential platform for studying emergent quantum many-body phenomena such as mediated interactions and unconventional pairing.
Here, we realize an equilibrium Bose-Fermi mixture in a bilayer electron system implemented in a WS$_{2}$/WSe$_{2}$ moiré heterobilayer with strong Coulomb coupling to a nearby moiré-free WSe$_{2}$ monolayer. 
Absent the fermionic component, the underlying bosonic phase manifests as a dipolar excitonic insulator. 
By injecting excess charges into it, we show that the bosonic phase forms a stable mixture with added electrons but abruptly collapses upon hole doping. 
We develop a microscopic model to explain the unusual asymmetric stability with respect to electron and hole doping. 
By studying the Bose-Fermi mixture via monitoring excitonic resonances from both layers, we demonstrate gate-tunability over a wide range in the boson/fermion density phase space, in excellent agreement with theoretical calculations.
Our results further the understanding of phases stabilized in moiré bilayer electron systems and demonstrate their potential for exploring the exotic properties of equilibrium Bose-Fermi mixtures.
\end{abstract}

\maketitle

\section*{Main}
\footnotetext{Amine.Ben-Mhenni@tum.de}
\stepcounter{footnote}
\footnotetext{JJ.Finley@tum.de}
\footnotetext{Michael.Knap@ph.tum.de}
\footnotetext{Nathan.Wilson@tum.de}
\setcounter{footnote}{1}
\footnotetext{These authors contributed equally to this work.}
\renewcommand{\thefootnote}{\arabic{footnote}}
       
Two-dimensional (2D) moiré materials are a leading platform to realize and study emergent quantum many-body phenomena \cite{kennes_moire_2021}, such as the recently discovered fractional quantum anomalous Hall effect \cite{park_observation_2023, kang_evidence_2024, lu_fractional_2024}.
In particular, semiconducting moiré bilayers based on transition metal dichalcogenides (TMDs) combine strong charge correlations with large exciton binding energies, enabling the study of a wide range of fermionic and bosonic collective phenomena \cite{wilson_excitons_2021, regan_emerging_2022, mak_semiconductor_2022, montblanch_layered_2023}.
Therein, the periodic moiré potential quenches the kinetic energy of charges, promoting strong charge correlations which result in many-body phases such as Mott insulators \cite{tang_simulation_2020, regan_mott_2020, xu_correlated_2020, shimazaki_strongly_2020}, generalized Wigner crystals \cite{regan_mott_2020, xu_correlated_2020}, and kinetic magnetism \cite{ciorciaro_kinetic_2023}.
It also enhances collective bosonic phenomena such as excitonic insulators \cite{ma_strongly_2021, zhang_correlated_2022, gu_dipolar_2022, chen_excitonic_2022, xiong_correlated_2023, lian_valley-polarized_2024, gao_excitonic_2024}, exciton density waves \cite{zeng_exciton_2023}, and dipole ladders \cite{park_dipole_2023}.

Outside of these purely bosonic or fermionic phases, mixtures of the two can lead to fundamentally distinct phases due to mutual interactions \cite{baranov_condensed_2012}.
A seminal example of this is BCS superconductivity which emerges due to phonon-mediated electron interactions \cite{bardeen_theory_1957}. 
Recent studies extended the realization of Bose-Fermi mixtures to TMD heterostructures, relying on optically pumped interlayer excitons as the underlying bosonic phase \cite{xiong_correlated_2023, tan_layer-dependent_2023, lian_valley-polarized_2024, gao_excitonic_2024, upadhyay_giant_2024}. 
These excitons are, however, relatively short-lived \cite{wilson_excitons_2021} and have significant excess energy due to nonresonant excitation, which could prevent the observation of fragile collective phenomena in Bose-Fermi mixtures.

\begin{figure*}[t]
\includegraphics[width=1\textwidth]{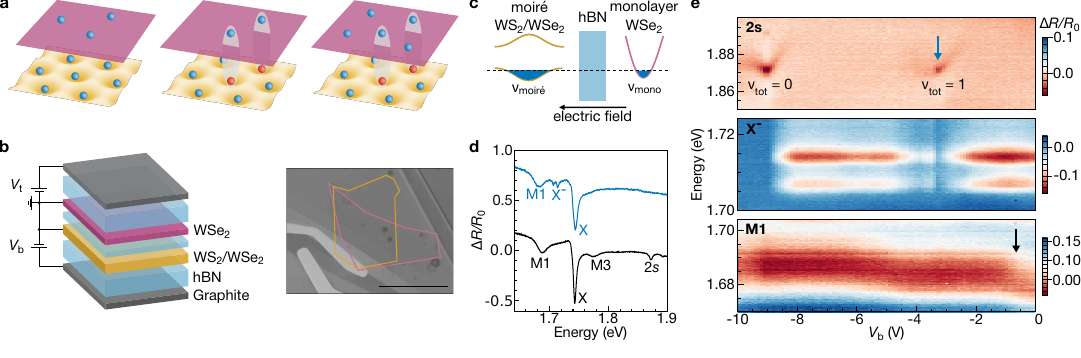}
\caption{\label{fig:device}\textbf{Device configuration and basic optical characterization of a bilayer electron system.}
\newline
\textbf{a,} Illustration of different correlated phases in a bilayer electron system with the lower layer being subject to a moiré superlattice. From left to right: Mott insulator, dipolar excitonic insulator, and Bose-Fermi mixture.
Electrons are depicted as blue spheres, and holes as red spheres. 
\textbf{b,} Schematic (left panel) and optical micrograph (right panel) of the device. 
In this bilayer electron system, an angle-aligned WS$_{2}$/WSe$_{2}$ moiré layer serves as the lower layer, and a moiré-free WSe$_{2}$ monolayer as the upper layer. The two layers are separated by an ultrathin ($\sim 1.3$ nm) hBN barrier, suppressing electron tunneling but allowing for strong Coulomb coupling. 
The scale bar is $20$ $\mu m$.
\textbf{c,} Schematic of the band alignment of the system. 
The band alignment can be tuned by applying an electric field such that the CB is either that of WS$_{2}$ in the heterobilayer or the WSe$_{2}$ monolayer. 
\textbf{d,} Reflection contrast spectra with both layers being in the charge-neutral (black) and negatively charged (blue) regime. The WSe$_{2}$ moiré excitons (M1 and M3) from the WS$_{2}$/WSe$_{2}$ moiré layer and the charge-neutral excitons (X and $2s$) and charged excitons (X$^{-}$) from the WSe$_{2}$ monolayer are labeled. 
\textbf{e,} Gate-dependent reflection contrast spectrum with an electron-doped WSe$_{2}$ monolayer ($V \mathrm{_{t}} = 6.25$ V).
}
\end{figure*}

Bilayer electron systems (Fig.~\ref{fig:device}a) offer an elegant route to overcome the limitation of short exciton lifetimes through the spontaneous formation of interlayer excitons via electrostatic gating, realizing an equilibrium population of stable bosons when charge tunneling between layers is suppressed \cite{eisenstein_boseeinstein_2004, ma_strongly_2021}. Dipolar excitonic insulators have been recently observed following this approach \cite{zhang_correlated_2022, gu_dipolar_2022}. Whether the same approach can be extended to realize equilibrium Bose-Fermi mixtures is an open challenge.

Here, we study a bilayer electron system in which one layer is a WS$_{2}$/WSe$_{2}$ moiré heterobilayer coupled by strong Coulomb interactions to a nearby WSe$_{2}$ monolayer, which serves as the second electron layer in the continuum. Relying on the strong charge correlations near the Mott transition in the moiré heterobilayer, we electrostatically inject bosonic dipolar excitons into the system. 
At charge balance, our system realizes a purely bosonic phase, a dipolar excitonic insulator, analogous to recent observations \cite{zhang_correlated_2022, gu_dipolar_2022}. 
By tuning the system away from charge balance, we observe an abrupt collapse of the state, provided the excess charges are holes. In strong contrast, we observe a stable equilibrium Bose-Fermi mixture when electrons are injected.
We develop a microscopic model explaining this strikingly asymmetric stability of the bosonic dipolar excitonic insulator phase to electron/hole doping. 
By studying excitonic resonances from both constituent layers, our findings show that the Bose-Fermi mixture is tunable over a wide range of boson/fermion densities in excellent agreement with our theoretical calculations. 
Our results demonstrate that bilayer electron systems can host highly tunable equilibrium Bose-Fermi mixtures and have great potential to explore fragile exotic phases that can emerge within them.

\subsection*{Charge correlations in a bilayer electron system with a moiré layer}

Our device consists of an angle-aligned WS$_{2}$/WSe$_{2}$ moiré layer with a triangular moiré superlattice with periodicity $a_\text{\moire} \sim 8$ nm, separated from a moiré-free WSe$_{2}$ monolayer by a $4$-layer hexagonal boron nitride (hBN) barrier (Fig.~\ref{fig:device}b and Methods).
The barrier is thick enough to suppress charge tunneling, separating the charges in the two layers \cite{ma_strongly_2021}.
At the same time, it is thin enough to maintain a strong interlayer Coulomb coupling comparable to the intersite Coulomb interaction in the moiré layer \cite{gu_dipolar_2022}.
The device is symmetrically dual-gated by $\sim 40$ nm hBN layers and graphite electrodes, allowing independent control of the electrochemical potential and the external electric field.
Specifically, we can tune the conduction band (CB) of the WSe$_{2}$ monolayer to be the global CB of the system (Fig.~\ref{fig:device}c), which otherwise originates from the moiré WS$_{2}$ in the absence of an external electric field.
Thus, we can independently control the electron filling factor of the moiré layer $\nu _\text{\moire}= n_\text{\moire}/n\mathrm{_{0}}$ and that of the WSe$_{2}$ monolayer 
$\nu \mathrm{_{mono}} = \nu \mathrm{_{tot}} - \nu _\text{\moire}$ by the top ($V \mathrm{_{t}}$) and bottom ($V\mathrm{_{b}}$) gate voltages.
Here, $\nu\mathrm{_{tot}}$ is the total filling factor of the system, $n_\text{\moire}$ the electron density in the moiré layer, and $n\mathrm{_{0}}$ the moiré density ($\sim 2 \times 10^{12}$ cm$^{-2}$).
For convenience, we express the charges in the WSe$_{2}$ monolayer as a filling factor $\nu \mathrm{_{mono}}$ per \moire unit cell.

Figure~\ref{fig:device}d shows reflection contrast spectra ($\Delta R / R_{0}$) of the bilayer electron system when both layers are charge-neutral (black) and electron-doped (blue).
The spectra are composed of excitonic resonances from the WSe$_{2}$ monolayer and from the moiré layer (See Extended Data Fig.~\ref{figext:excitons}).
The WSe$_{2}$ monolayer hosts the neutral exciton (X), which turns into the repulsive polaron upon charge doping, and the negatively charged exchange-split trions (X$^{-}$).
X$^{-}$ and the repulsive polaron are sensitive to the presence of free electrons in the WSe$_{2}$ monolayer. Changes in the energy and strength of X and X$^{-}$ reflect changes in the free carrier density, or compressibility, including changes stemming from charge correlations \cite{gu_dipolar_2022, cai_signatures_2023}.

The monolayer also hosts the first excited state of X ($2s$), which has a large Bohr radius of several nanometers \cite{stier_magnetooptics_2018}, larger than the separation between the two layers of the bilayer electron system. 
Since its electric field permeates both layers of the system, it is sensitive to the overall compressibility of charges therein \cite{xu_correlated_2020}, losing oscillator strength and blueshifting in the presence of free charges due to increased screening.
When the WSe$_{2}$ monolayer is charge neutral (itself in a trivial incompressible state), the $2s$ exciton can detect incompressible states at fractional and integer fillings of the nearby moiré layer superlattice (See Extended Data Figs. \ref{figext:neutral}, \ref{figext:zoo}). 
Moreover, even when the individual layers are charge compressible, the $2s$ exciton can sense an \textit{overall incompressible state} of the system arising from interlayer charge correlations, as in the observed dipolar excitonic insulator \cite{zhang_correlated_2022, gu_dipolar_2022}.
Nevertheless, the $2s$ exciton quickly loses oscillator strength upon injection of free charge carriers into the monolayer, restricting its utility to sense charge correlations.

The WSe$_{2}$ in the moiré layer hosts moiré excitons M$1$, M$2$, and M$3$ instead of free excitons (See Extended Data Fig. \ref{figext:excitons}), owing to the strong moiré potential that generates flat exciton minibands \cite{jin_observation_2019, wu_topological_2017}.
The spectral response of all the moiré excitons is affected by the charge ordering in the moiré layer \cite{xu_correlated_2020, liu_excitonic_2021, tang_simulation_2020}. In particular, the moiré excitons can detect changes in screening due to charge correlation or abrupt jumps in the chemical potential, i.e. when the Fermi level crosses a charge gap, including correlation gaps. 
Among the moiré excitons, M$1$ has the strongest optical contrast over a large range of electron density and is spectrally separated from the resonances of the WSe$_{2}$ monolayer. 
Therefore, we will focus our analysis on M1 and use it to probe the charge ordering in the moiré layer.

We now analyze the distinct spectral features in reflection contrast while sweeping $V\mathrm{_{b}}$ at $V\mathrm{_{t}} = 6.25 $ V (Fig.~\ref{fig:device}e). 
Across most of the gate range, the band alignment in the bilayer is approximately as depicted in Fig. ~\ref{fig:device}c, with the Fermi level being tuned through the lower electron Hubbard band of the moiré layer and sitting within the CB of the monolayer.
At $V\mathrm{_{b}} \sim -9$ V, a trivial band insulator is realized in both layers ($\nu \mathrm{_{mono}} = 0$, $\nu _\text{\moire} = 0$), resulting in a strong $2s$ exciton resurgence and an enhanced M1 contrast.
As $V\mathrm{_{b}}$ increases, the number of electrons distributed across both layers increases.
This is made clear by the presence of X$^{-}$ in the monolayer across most of the plotted gate range. At the same time, filling of subsequent electron Hubbard bands in the moiré layer causes the observed energy jumps in M1 \cite{liu_excitonic_2021}. 

In addition to the trivial band insulator at $V\mathrm{_{b}} \sim -9$ V, the redshift and sudden change in contrast of M1 at $V\mathrm{_{b}} \sim -0.5$ V indicates the presence of a Mott insulator in the moiré layer (black arrow in Fig.~\ref{fig:device}e, bottom panel). 
Notably, this does not coincide with the incompressible state detected by the $2s$ exciton at $V\mathrm{_{b}} \sim -3.5$ V (blue arrow in Fig.~\ref{fig:device}e, top panel). 
At this $V\mathrm{_{b}}$, the Mott insulator in the moiré layer has been doped by holes.
Furthermore, in the vicinity of the incompressible state, the monolayer is populated by free electrons, which give rise to the strong X$^{-}$ signal shown in Fig.~\ref{fig:device}e (middle panel). 
The electrons in the monolayer and holes in the moiré layer interact attractively and can bind to form bosonic dipolar interlayer excitons. 
When the number of electrons and holes is equal, exciton formation depletes both layers of free charges \cite{cai_signatures_2023}, causing the observed disappearance of the X$^{-}$ signal and producing the incompressible dipolar insulating phase \cite{zhang_correlated_2022, gu_dipolar_2022} detected by the $2s$ exciton (See Extended Data Fig. \ref{figext:hole_side} for the dipolar excitonic insulator with opposite polarity at $\nu \mathrm{_{tot}} = -1$). 

The existence of the dipolar insulator as a distinct state is further corroborated by a different melting temperature of $\sim 150$ K compared to at least $210$ K for the Mott insulating state in the moiré layer (See Extended data Fig.~\ref{figext:temperature}).
In the following section, we will show that the dipolar insulator branches and diverges from the Mott insulating state upon electron doping of the WSe$_{2}$ monolayer.

\subsection*{Stability of the dipolar excitonic phase to charge doping}

\begin{figure*}
\includegraphics[width=1\textwidth]{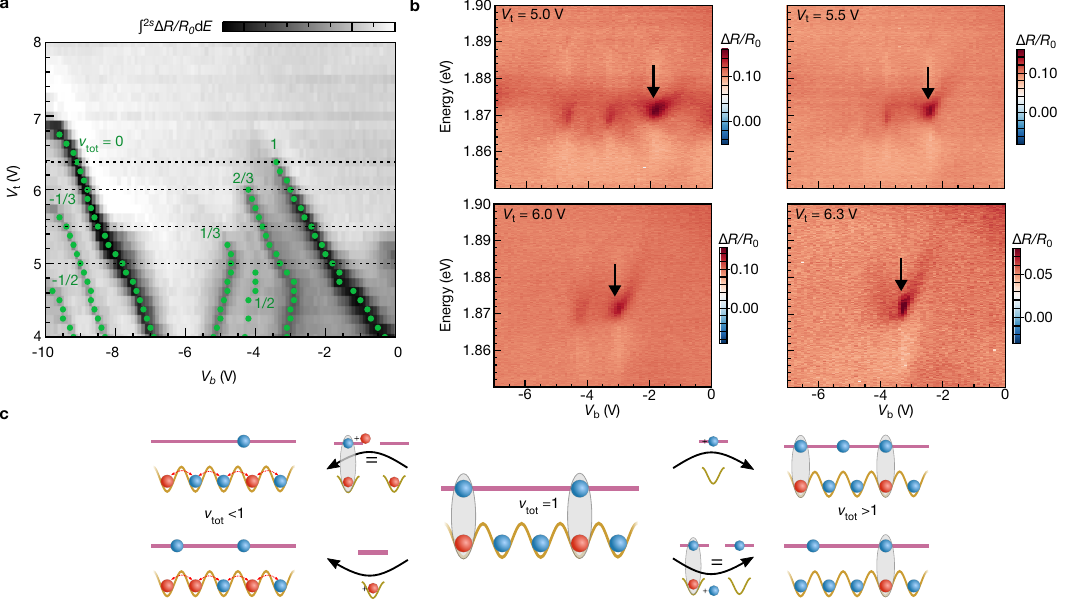}
\caption{\label{fig:asymmetry}\textbf{Asymmetric stability of the dipolar phase subject to electron/hole doping.}
\newline
\textbf{a}, Electrostatic phase diagram of the $2s$ exciton. 
The emergence of the $2s$ resonance signal corresponding to the incompressible states of the bilayer electron system is detected via peak prominence and overlayed in green. 
\textbf{b,} Gate-dependent ($V\mathrm{_{b}}$) reflection contrast spectra of the $2s$ exciton at $V\mathrm{_{t}}$ = $5$ V, $5.5$ V, $6$ V, and $6.3$ V exhibiting a gradual transition from a symmetric to an unusual asymmetric behavior in $V\mathrm{_{b}}$. 
\textbf{c,} Starting from the dipolar excitonic insulator with one free-layer electron per one hole in the \moire layer (middle), electron doping (right) effectively introduces more electrons in the free layer, coexisting with the dipolar excitonic insulator.
Hole doping (left) destroys the Mott insulating state in the \moire layer and, with it, the dipolar excitonic insulator. 
Both electron and hole doping do not depend on the layer in which the charges are inserted because of the possible destruction of dipolar excitons (See insets between the main panels). Electrons are depicted as blue spheres, and holes as red spheres. 
}
\end{figure*}

Having established the presence of the charge-balanced dipolar excitonic insulator, we now intentionally create a charge imbalance between the layers to study the effect of doping the phase with excess electrons and holes.

To this end, we determine the electrostatic phase diagram ($V\mathrm{_{b}}$, $V\mathrm{_{t}}$) obtained by integrating over the $2s$ resonance (Fig.~\ref{fig:asymmetry}a).
There we trace the resurgence of the $2s$ exciton via maximum peak prominence, indicated by the green dots.
When no charges are present in the WSe$_{2}$ monolayer ($\nu \mathrm{_{tot}} = \nu _\text{\moire}$), the lines with strong $2s$ contrast are consistent with the correlated states at fractional fillings of the moiré superlattice: the trivial band insulator at $\nu _\text{\moire}= 0$, the electron generalized Wigner crystals at $\nu _\text{\moire}= 1/3$ and $2/3$, the stripe phase at $\nu _\text{\moire}= 1/2$, and finally, the Mott insulator at $\nu _\text{\moire}= 1$ \cite{xu_correlated_2020}.
As discussed in the previous section, the Mott insulator at $\nu \mathrm{_{tot}} = 1$ transforms into the dipolar excitonic phase upon redistribution of the charges from the moiré layer to the WSe$_{2}$ monolayer, i.e., upon increasing $V_\text{t}$.

\begin{figure*}[t]
\includegraphics[width=1\textwidth]{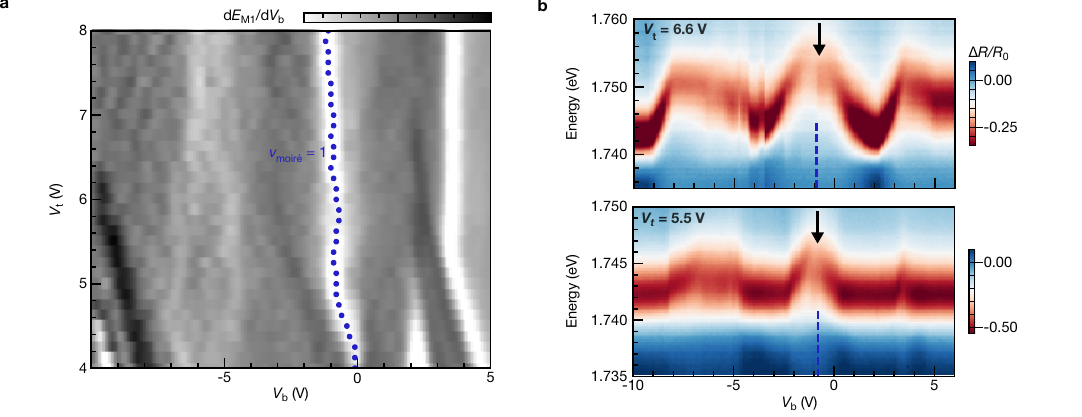}
\caption{\label{fig:mott_phase}\textbf{Identification of the Mott insulator throughout the electrostatic phase diagram.}
\newline
\textbf{a,} Electrostatic phase diagram of M1, obtained from the derivative of its energy with respect to $V\mathrm{_{b}}$. 
The blue dotted curve indicates the energy redshifting of M1 corresponding to the Mott insulating phase in the moiré layer. 
\textbf{b,} Gate-dependent ($V\mathrm{_{b}}$) reflection contrast spectrum of the polaron in the monolayer at two different values of $V\mathrm{_{t}} = 5.5$ V (lower panel), and $6.6$ V (upper panel).
The black arrows indicate resurgences of the polaron, coinciding with the redshifting of M1.
The dashed lines indicate the energy redshifting of M1 from (a).
}
\end{figure*}

In the following, we focus on the behavior of the $2s$ resonance in the vicinity of the excitonic dipolar insulator.
The $2s$ exciton spectroscopic response obtained via reflection contrast $V\mathrm{_{b}}$ sweeps at $V\mathrm{_{t}} = 5$ V, $5.5$ V, $6$ V, and $6.3$ V are shown in Fig.~\ref{fig:asymmetry}b. 
The arrows indicate the voltage that achieves the charge-balanced dipolar excitonic state ($\nu \mathrm{_{tot}} = 1$). 
To the left of the arrow, holes are added to the system, and to the right, electrons are added. 
As $V\mathrm{_{t}}$ increases, $\nu \mathrm{_{mono}}$ and, therefore, the density $\nu\mathrm{_{dipole}}$ of the dipolar excitons at charge balance increases.
At lower $V\mathrm{_{t}}$ (low $\nu\mathrm{_{dipole}}$), the $2s$ exciton loses oscillator strength and blueshifts continuously, as tuning $V\mathrm{_{b}}$ away from charge balance injects holes or electrons to the charge-balanced phase. 
Remarkably, as $V\mathrm{_{t}}$ increases towards $6.25$ V (and along with it $\nu\mathrm{_{dipole}}$), the evolution of the $2s$ exciton signal becomes strongly asymmetric with respect to electron/hole doping. 
It dims and blueshifts gradually for electron doping but disappears suddenly upon hole doping, suggesting an abrupt change of the quantum state. 

We now develop a microscopic picture to elucidate the origin of the electron/hole doping asymmetry around the bosonic phase and sketch the bilayer electron system around the dipolar excitonic insulator (Fig.~\ref{fig:asymmetry}c).
Starting from the charge-balanced case (middle panel), injecting an excess electron in either layer effectively amounts to injecting an electron in the WSe$_{2}$ monolayer since an excess electron in the moiré layer recombines with the hole of a dipolar exciton, freeing an electron in the WSe$_{2}$ monolayer (right panel).
While this additional free electron in the WSe$_{2}$ monolayer increases the screening in the system, it does not directly undermine the stability of the dipolar excitonic insulator.
This picture is consistent with the observed spectroscopic behavior of the $2s$ exciton increase of the free electrons in the WSe$_{2}$ monolayer results in the gradual blueshifting and fading away of the $2s$ resonance.
Remarkably, the situation is completely different when holes are injected instead.
Unlike the electron case, injecting an excess hole into either layer of the bilayer electron system results in an excess hole in the moiré layer. This is because an excess hole in the WSe$_{2}$ monolayer simply destroys a dipolar exciton and sets a hole in the moiré layer free (left panel).
The excess hole can tunnel within the moiré layer, causing the abrupt collapse of the dipolar phase.
Here, the increase in the dielectric screening by the moiré layer and the sudden jump in the chemical potential in the WSe$_{2}$ monolayer result in an abrupt fading away of the $2s$ resonance, consistent with our observations.
Moreover, analogous conclusions apply to the dipolar excitonic insulator with opposite polarity (at $\nu \mathrm{_{tot}} = -1$) but with mirrored asymmetric stability behavior in $V\mathrm{_{b}}$, that is, the dipolar state is stable to hole doping but collapses upon excess electron injection (See Extended Data Fig.~\ref{figext:hole_side}).

\subsection*{Extent of the Bose-Fermi mixture}

\begin{figure*}[t]
\includegraphics[width=1\textwidth]{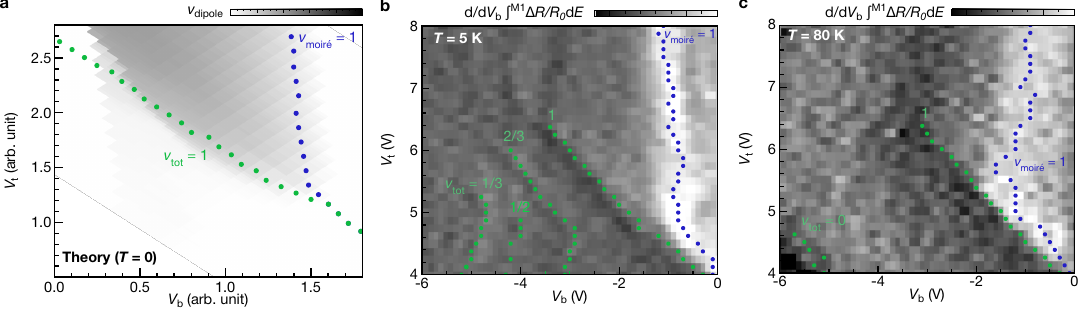}
\caption{\label{fig:theoryandsummary}\textbf{Bose-Fermi mixture at different temperatures.}
\newline
\textbf{a,} Theoretical calculations of the bilayer electron system in the ground state ($T=0$).
The dipole density is finite in the region delimited by the dipolar insulting phases $\nu_\text{tot}=1$ and the Mott insulator $\nu_\text{\moire}=1$.
\textbf{b,} Electrostatic phase diagram of M1 at $5$ K.
Here, M1 was integrated over the energy range $1.680$ eV to $1.690$ eV, and the derivative with respect to $V\mathrm{_{b}}$ was taken.
The $2s$ resonance signal and the energy redshifting of M1 are overlayed in green and blue, respectively. 
\textbf{c,} Electrostatic phase diagram of M1 at $80$ K.
Here, M1 was integrated over the energy range $1.670$ eV to $1.685$ eV, then the derivative with respect to $V\mathrm{_{b}}$ was taken.
The $2s$ resonance signal and the energy redshifting of M1 are overlayed in green and blue, respectively. 
}
\end{figure*}

Our results thus indicate that the bosonic phase, while fragile to hole doping, is generally robust to electron doping.
This suggests that the incompressible charge-balanced dipolar excitonic insulator is a \textit{boundary} of a larger region consisting of a mixture of bosonic dipolar excitons and fermionic free electrons residing in the monolayer WSe$_{2}$.

We now study the extent of the mixed phase in terms of dipole and charge densities. 
Even when electrons are present in the WSe$_{2}$ monolayer, if the moiré layer is in the Mott insulating phase, there will be no available holes in the lower Hubbard band that could bind with them to form dipolar excitons.
Therefore, identifying the Mott insulating state across the electrostatic phase diagram amounts to finding the maximum extent up to which the Bose-Fermi mixture could possibly exist.

To identify the Mott insulating phase, we conduct a detailed analysis of the optical response of M1 \cite{tang_simulation_2020}, extracting the electrostatic phase diagram from the energy shift of the M1 resonance (Fig.~\ref{fig:mott_phase}a).
We fit M1 with a dispersive Lorentzian profile \cite{shimazaki_strongly_2020} at each point of the phase diagram, extracting the center energy from the fit and taking its first derivative with respect to $V\mathrm{_{b}}$.
The overlaid blue dotted line tracks the derivative's minimum. 
When the WSe$_{2}$ monolayer is free of charges, this line coincides with the Mott insulating state ($\nu \mathrm{_{tot}} = \nu _\text{\moire}= 1$).

To further corroborate our identification of the Mott insulating state in the moiré layer, we turn to study the repulsive polaron residing in the WSe$_{2}$ monolayer, which is a neutral exciton that dressed by the electrons in the Fermi sea \cite{sidler_fermi_2017}. 
This polaron, which is larger than a neutral exciton, persists over large density ranges and exhibits charge density-dependent energy shift and intensity \cite{sidler_fermi_2017, xu_creation_2021, ben_mhenni_breakdown_2024}.
It is, therefore, sensitive to variations of the chemical potential of the WSe$_{2}$ monolayer and of the dielectric function of both constituent layers of the system.
Figure ~\ref{fig:mott_phase}b shows the polaron spectroscopic response measured via reflection contrast $V\mathrm{_{b}}$ sweeps at $V\mathrm{_{t}} \sim 5.5 $ V (lower panel) and $\sim 6.6 $ V (upper panel).
The polaron undergoes a series of resurgences comparable in appearance to the behavior of the $2s$ exciton when sensing correlated phases of Hubbard systems \cite{xu_correlated_2020}.
This behavior becomes more pronounced once the electron density in the WSe$_{2}$ monolayer is sufficiently high (top panel).
Therefore, the resurgence of the polaron reflects a sudden change in compressibility occurring within the moiré layer due to the sensitivity of the polaron to changes in dielectric screening \cite{ben_mhenni_breakdown_2024}.
Strikingly, it matches well with the extracted position of the redshifting of M1 (blue dashed lines), indicating a common origin interpreted as the Mott insulating state in the moiré layer.
When the moiré layer is in the Mott state and the WSe$_{2}$ monolayer is electron-doped ($\nu _\text{\moire}= 1$ and $\nu \mathrm{_{mono}} > 0$), the two layers are expected to be uncorrelated and the interlayer excitons cannot form \cite{tan_layer-dependent_2023}. 
In this case, the presence of the free electrons in the WSe$_{2}$ monolayer will also screen the long-range Coulomb interactions within the moiré layer.
When the long-range Coulomb interactions within the moiré layer are completely suppressed, the many-body physics of this layer is described by the simple Hubbard model instead of its extended version \cite{tang_evidence_2023}.

Having delimited the possible boundaries of the dipolar excitonic phase, we now study the phase diagram theoretically to gain further insights into the stability of the mixture of interlayer excitons and charge dopants.
We describe the bilayer electron system with a purely electronic model Hamiltonian to further investigate the robust coexistence of emergent bosonic and fermionic particles (Methods).
In contrast to previous models of continuous electron-electron or electron-hole bilayers~\cite{Zhu1995, eisenstein_boseeinstein_2004, Wu2015}, we consider the effect of the superlattice in the moiré layer.
Therefore, our model consists of heavy electrons on a triangular lattice in the moiré layer and light electrons in the WSe$_{2}$ monolayer.
Direct tunneling between the layers is prohibited by the hBN spacer, but the layers are coupled via Coulomb interactions.
We also include strong intralayer interactions, according to parameter estimates from band theory calculations \cite{wu_hubbard_2018}.
To solve the interacting Hamiltonian, we decouple the Coulomb interactions into density mean fields.
In that way, we directly obtain the filling in each layer and the number of dipoles from self-consistent solutions that minimize the mean-field energy.
In the model, a displacement field tunes the chemical potential difference between the layers, and we simulate a range of fixed total fillings.
We transform the obtained phase diagrams to the experimentally accessible ones, that is, in terms of the gate voltages $V_\text{t}$ and $V_\text{b}$, with a parallel plate capacitor model \cite{tan_layer-dependent_2023}.

Using our model, we determine the average dipole density $\nu_\text{dipole}$ as a function of the gate voltages (Fig.~\ref{fig:theoryandsummary}a).
Our approach predicts an extended region of the phase diagram where a finite density of dipoles are stabilized in the purely electronic model.
This region occurs in the entire area between the line of total filling $\nu_\text{tot} = 1$ and the Mott insulator $\nu_\text{\moire} = 1$.
Importantly, we observe the same asymmetry as in the experiment when doping either holes or electrons into the dipolar excitonic insulator:
dipoles survive upon electron doping, leading to a robust Bose-Fermi mixture, while hole doping immediately destroys the dipolar state.

By combining the previous findings, we form our understanding of the system's phase diagram.
We extract the electrostatic phase diagram at 5 K by integrating the signal of M1 over the energy range $1.68$ eV to $1.69$ eV and taking the derivative with respect to $V\mathrm{_{b}}$ (Fig.~\ref{fig:theoryandsummary}b).
The curve corresponding to $\nu _\text{\moire}= 1$, obtained via the fitting of M1 (Fig.~\ref{fig:mott_phase}a) is overlaid as a dotted blue line, while the curve corresponding to $\nu \mathrm{_{tot}} = 1$, obtained via the analysis of the $2s$ exciton (Fig.~\ref{fig:asymmetry}a) is overlaid as a dotted green line.
Both overlaid curves match with an increase and decrease in M1's reflection contrast when decreasing $V\mathrm{_{b}}$ at a constant $V\mathrm{_{t}}$, respectively.
Furthermore, M1 maintains a rather constant reflection contrast between the two curves without sudden changes, consistent with a continuously tuned Bose-Fermi mixture in this region.
Both curves merge together into the Mott insulating state when the WSe$_{2}$ monolayer becomes charge neutral at $V\mathrm{_{t}} \approx 4.7$ V.

We now study the stability of the Bose-Fermi mixture at higher temperatures. To this end, we determine the electrostatic phase diagram of M1 at $80$ K (Fig.~\ref{fig:theoryandsummary}c) using the same analysis approach as for $5$ K. 
By identifying the dipolar excitonic insulator phase (at $\nu_\text{tot} = 1$) via the $2s$ resonance and the Mott insulator ($\nu_\text{\moire} = 1$) via M1, at $80$ K we observe an analogous behavior as for $5$ K, but with a smaller region of stability in the $V\mathrm{_{t}}$ and $V\mathrm{_{b}}$ phase space. 
This indicates that in our setting the Bose-Fermi mixture persists up to at least $80$ K.
When further increasing the temperature beyond $80$ K, the analysis of M1 across the electrostatic phase diagram becomes inconclusive due to broadening and weaker contrast. The electrostatic phase diagram of the $2s$ exciton, however, indicates that the dipolar excitonic insulator persists up to approximately 150 K, see Extended Data Figure~\ref{figext:temperature}.

\subsection*{\label{sec:main:outlook}Outlook}

Our results reveal the existence of a correlated phase consisting of spontaneously formed bosonic interlayer excitons and itinerant electrons, realizing a stable and highly tunable Bose-Fermi mixture. 
Because the excitons in the mixture are electrically injected and their recombination is suppressed by the hBN barrier layer, they are expected to be long-lived compared to optically injected excitons, 
which allows them to thermalize with the electrons and the lattice. 
Due to their fixed out-of-plane dipole moment, the excitons experience long-range dipole-dipole interactions. This could give rise to density-wave instabilities of the excitons at fractional fillings which would be interesting to explore in future work. 
Moreover, one could explore whether Bose-Fermi mixtures can be stabilized in the vicinity of generalized Wigner crystals, that are stabilized due to the Coulomb interactions in the moiré layer.

The combination of stability, density tunability, long lifetime, large effective masses inherited from the moire layer, and potential for strong long-range interactions makes this an attractive system to explore correlation effects. 
Bose-Fermi mixtures are predicted to exhibit a rich phase diagram of correlated states and phenomena, including supersolids~\cite{buchler_supersolid_2003, matuszewski_exciton_supersolid_2012}, unconventional transport~\cite{yan_collective_2024}, and topological superconductivity~\cite{zerba_realizing_2024, von_milczewski_superconductivity_2024}, with interactions tunable through the population densities and other external means such as solid-state Feshbach resonances~\cite{schwartz_electrically_2021, kuhlenkamp_tunable_2022, zerba_tuning_2024}.
For future studies, it would also be intriguing to investigate the spin order in these systems. 
Previous works suggest weakly antiferromagnetic correlations for the dipolar excitonic insulator~\cite{gu_dipolar_2022}, but additional charges may also kinetically induce a ferromagnetic order~\cite{ciorciaro_kinetic_2023, Yang2024}.
The new level of stability and control of these moiré bilayer electron systems could thus enable to study several exotic properties of equilibrium Bose-Fermi mixtures. 

\section*{\label{sec:Methods}Methods}

\subsection*{\label{sec:methods:sample}Sample preparation}
All TMD, graphite, and hBN flakes were mechanically exfoliated from bulk crystals on $70$ nm SiO$_{2}$ substrates. 
Homogeneous flakes were selected based on their optical contrast, shape, and cleanliness.
The thickness of the $4$-layer hBN spacer and other hBN layers was confirmed by atomic force microscope measurements.
The device was assembled via the dry-transfer technique using polycarbonate (PC) films. 
The picking up of flakes was done at $120 ^{\circ}$ C.
After assembly of the heterostructure on the stamp, the interfaces were cleaned by repeatedly contacting and picking up the heterostructure at $155 ^{\circ}$ C, which mechanically squeezes out trapped bubbles between the constituent layers.
The sample was subsequently submerged in chloroform ($30$ min) and IPA ($60$ min) to remove the polymer before contacting the respective layers using standard maskless optical lithography and subsequent electron beam evaporation of Cr/Au $5/100$ nm electrodes.

\subsection*{\label{sec:methods:spectroscopy}Optical spectroscopy}

The optical measurements were performed in a close-cycle optical cryostat in reflection geometry (Attocube, attoDRY800).
For the reflection contrast measurements, thermal light from a tungsten halogen light source was focused onto the sample using a $40$X objective with a numerical aperture of $
0.75$, yielding an excitation spot size of around $1$ $\mu$m. 
A pinhole was used as a spatial filter to obtain a diffraction-limited collection spot. The collected light was dispersed using a $500$ mm focal length spectrometer and detected on a back-illuminated CCD sensor array.
The gates were controlled using a source-measure unit with monitored leakage current.
Unless otherwise specified, all measurements presented here were performed at $5$ K.
\newline

\subsection*{Theoretical model and mean-field solution}

We describe the multilayer heterostructure with free electrons in the monolayer WSe$_2$ and triangular-lattice electrons in the WSe$_2$/WS$_2$ \moire layer as
\begin{subequations}
\begin{align}
    H =& H_\text{\moire} + H_\text{mono} + H_\text{inter} \label{eq:full_hamiltonian} \\
    H_\text{\moire} =& -t^f \sum_{\langle i, j \rangle} \left( \fd{i} \fo{j} + \text{h.c.} \right) \nonumber - \mu^f \sum_i \fd{i}\fo{i} \\
    &+ V \sum_{\langle i, j \rangle} n^f_i n^f_j \\
    H_\text{mono} =& -t^c \sum_{\langle n, m \rangle} \left( \cd{n} \co{m} + \text{h.c.} \right) \nonumber - \mu^c \sum_n \cd{n}\co{n} \\
    &+ \frac{1}{2} \sum_{n, m, k, l} U^c_{n, m, k, l} \cd{n} \cd{m} \co{k} \co{l} \\
    H_\text{inter} =& \frac{1}{2} \sum_{i, j, n, m} U^{c}_{i, j, n, m} \cd{n} \fd{i} \fo{j} \co{m} .
\end{align}
\end{subequations}
The first term $H_\text{\moire}$ describes electrons, created (annihilated) by $\fd{i}$ ($\fo{i}$), in the \moire layer with hopping strength $t^f$ and nearest-neighbor interactions $V$. We estimate $t^f\approx 1.5\,$meV and $V\approx 10\,$meV for a \moire period of $a_\text{\moire}\approx 10\,$nm from band structure calculations~\cite{wu_hubbard_2018}. Since we focus on charge-ordered states with filling $\langle n \rangle \leq 1$ in each layer, we neglect the spin degree of freedom and work with a single-band Hamiltonian. The Hamiltonian $H_\text{mono}$ characterizes the electrons in the monolayer, which we associate with $\cd{n}$ ($\co{n}$) for the creation (annihilation) operators. For practical purposes, we discretize the continuum, but with a smaller unit cell of 1/9th of the \moire unit cell. Accordingly, we take $t^c \sim a_\text{mono}^{-2}/2m_c$. We are interested in the limit of small fillings in the monolayer $\langle n^c \rangle \ll \langle n^f \rangle$. Thus, any lattice effects and the precise form of the interactions do not change our results. 
Monolayer intralayer interactions and interlayer interactions are described with a screened Coulomb potential $U^c(r) = e^2/\varepsilon\,\left(1/r-1/{\sqrt{r^2+4d^2}} \right)$, where $d\approx 40$\,nm is the distance to the metallic gates and $\varepsilon\approx5$ to include the dielectric environment of hBN. We simplify the interlayer interactions to only include onsite interactions $H_\text{inter} \sim \cd{i} \co{i} \fd{i} \fo{i}$.

We perform a mean-field decoupling of the interactions by introducing the fields $\langle n^f_i \rangle$ and $\langle n^c_n \rangle$ for the intralayer interactions. For the interlayer interactions, we additionally couple the mixed densities $\langle \fd{i} \co{i} \rangle$.
We numerically solve the mean-field Hamiltonian self-consistently. For better convergence, we impose inversion and translation symmetries on our solution.
Furthermore, we find better convergence of the mean-field parameters when specifying the displacement field $F$ and total filling $\nu_\text{tot}$ instead of the chemical potentials $\mu^f$ and $\mu^c$. 

To relate the mean-field solutions to the experimental observations, we compute the number of excitons in the system.
Consider, for example, a single site on both layers. A general state can be written as $\ket{\psi}=\alpha \ket{0_f 0_c}+\beta \ket{1_f 0_c} + \gamma \ket{0_f 1_c} + \delta \ket{1_c 1_f}$. 
Having the Mott insulator $\ket{1_f 0_c}$ as ``vacuum'' state, the dipolar states we are interested in are $\ket{0_f 1_c}$. We thus define the dipole density $\nu_\text{dipole}= |\gamma|^2$.

Moreover, the experiment probes correlated states with the 2s exciton, which can be screened by free charge carriers in its surroundings. Hence, we also extract the number of free holes in the \moire layer Mott insulator and free electrons in the monolayer that are not bound as excitons, $\nu_\text{charges}=\underbrace{\langle n^c \rangle-\nu_\text{dipole}}_{\text{from monolayer}} + \underbrace{1- \langle n^f \rangle -\nu_\text{dipole}}_{\text{from \moire layer}}$.
While the full screening process is certainly more complicated and depends on microscopic details, this approximation will give us a reasonable intuition of the parameters where we can expect more screening.

To relate the displacement field and filling factor from the simulation to the top and bottom gate voltage of the experimental setup, we employ the parallel plate capacitor model~\cite{ tan_layer-dependent_2023}
\begin{subequations}
\begin{align}
    V_t &= \frac{1}{2C_g}(n_0 \nu_\text{tot}+ C_\text{inter} F) \\
    V_b &= \frac{1}{2C_g}(n_0 \nu_\text{tot}- C_\text{inter} F),
\end{align}
\end{subequations}
with the geometric capacitance $C_{g/\text{inter}} = \varepsilon A/d_{g/\text{inter}}$ for the distance of the layers to the gates and between the layers, respectively.

Our theoretical model can give insights into the underlying physical mechanisms and reproduce qualitatively the electrostatic phase diagram, see Fig.~\ref{fig:theoryandsummary} in the main text, with an extended region of finite dipole density. Moreover, in this region, a Bose-Fermi mixture is realized, which we identify by computing the number of electrons in the monolayer that is not bound in dipoles $\nu_\text{mixture}=\nu_\text{dipoles} - \nu_\text{mono}$. The number of free fermions increases when tuning $V_b$ in Extended Data Fig.~\ref{figext:theory}a from $\nu_\text{tot}=1$ (green) to the Mott insulator at $\nu_\text{\moire}=1$ (blue). While the individual charge densities in the mono layer and the \moire layer are continuously evolving, we also observe a reduced number of free charges $\nu_\text{charges}$ in the regime of the Bose-Fermi mixture, Extended Data Fig.~\ref{figext:theory}b, c, and d. This is because a sizeable portion of charges are bound in excitons. The reduced number of free charges gives rise to less screening and allows us to use the 2s exciton as a robust probe of the Bose-Fermi mixture.

\section*{\label{sec:data_availability}Data availability}
The datasets generated and analyzed during the current study are available from the corresponding authors upon reasonable request.

\bibliography{references}

\section*{\label{sec:acknowledgements}Acknowledgments}
We thank Xiaodong Xu and Clemens Kuhlenkamp for fruitful discussions.
A.B.M acknowledges funding from the International Max Planck Research School for Quantum Science and Technology (IMPRS-QST).
We gratefully acknowledge funding from the Deutsche Forschungsgemeinschaft (DFG, German Research Foundation) via Germany’s Excellence Strategy (MCQST, EXC-2111/390814868), large equipment grants INST95-1642-1 and INST 95/1719-1, SPP-2244 and the European Research Council (ERC) under the European Union’s Horizon 2020 research and innovation programme (grant agreement No. 851161). We also acknowledge the Munich Quantum Valley, which is supported by the Bavarian state government with funds from the Hightech Agenda Bayern Plus.
We thank the Nanosystems Initiative Munich (NIM), funded by the German Excellence Initiative and the Leibniz Supercomputing Centre, for access to their computational resources for the theoretical calculations.
A.O.P. acknowledges funding by the Bavarian Hightech Agenda within the Munich Quantum Valley doctoral fellowship program.
A.D. acknowledges funding from the European Union’s Horizon 2020 research and innovation programme under the Marie Skłodowska-Curie (grant agreement No. 101111251).
K.W. and T.T. acknowledge support from the JSPS KAKENHI (Grant Numbers 20H00354 and 23H02052) and World Premier International Research Center Initiative (WPI), MEXT, Japan.
S.A.T. acknowledges primary support from DOE-SC0020653 (materials synthesis), Applied Materials Inc., NSF CBET 2330110, DMR 2111812, DMR 2206987, and CMMI 2129412. S.A.T. also acknowledges support from Lawrence Semiconductor Labs.

\section*{\label{sec:contributions}Author contributions}
A.B.M. and N.P.W. conceived and managed the research. A.B.M., and A.O.P. fabricated the device. A.B.M., M.J.M., A.O.P., A.D., and N.P.W. performed the optical measurements. M.J.M. tailored software for the measurements. A.B.M. and N.P.W. analyzed the results in consultation with J.J.F. W.K. and M.K. developed the models and performed the calculations. S.T. grew WS$_{2}$ bulk crystals. K.W. and T.T. grew bulk hBN crystals. A.B.M, W.K., J.J.F, M.K, and N.P.W. prepared the manuscript in consultation with all other authors.
\newline

\section*{\label{sec:interests}Competing interests}
The authors declare no competing interests.

\renewcommand{\figurename}{\textbf{Extended Data Figure}}  
\setcounter{figure}{0}    

\section*{\label{sec:si}Extended data}

\begin{figure*}
\includegraphics[width=1\textwidth]{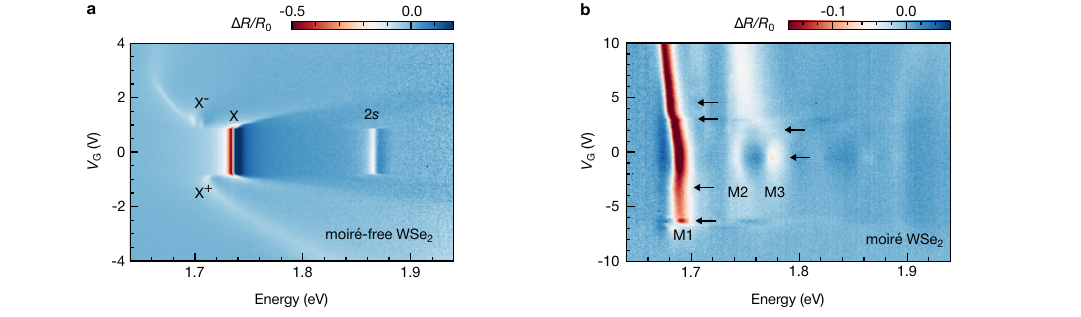}
\caption{\label{figext:excitons}\textbf{Gate-dependent WSe$_{2}$ spectrum: emergence of moiré excitons.}
\newline
Gate-dependent reflection contrast spectrum of a WSe$_{2}$ monolayer device (\textbf{a}) and of the WS$_{2}$/WSe$_{2}$ heterobilayer region of the device of this study (\textbf{b}). 
Instead of the moiré-free excitons (X, X$^{-}$, X$^{+}$, and $2s$), WSe$_{2}$ of the moiré heterobilayer exhibits moiré excitons M$1$, M$2$, and M$3$. 
These moiré excitons exhibit energy shifts and amplitude modulations corresponding to jumps in the chemical potential stemming from charge correlations (black arrows).
}
\end{figure*}

\begin{figure*}
\includegraphics[width=1\textwidth]{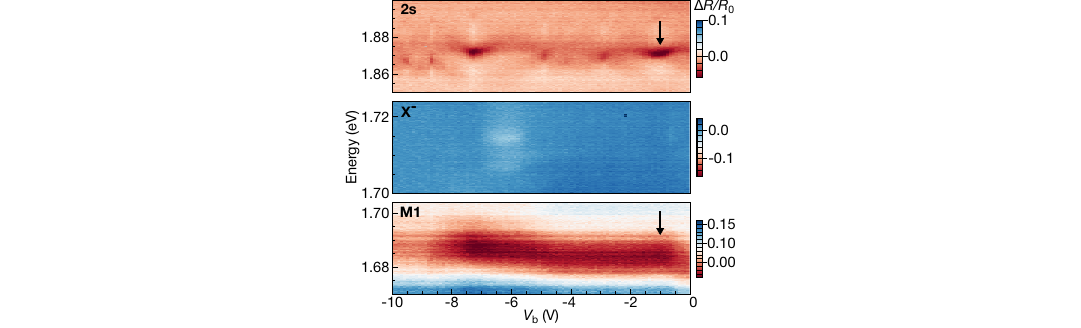}
\caption{\label{figext:neutral}\textbf{Bilayer electron system with a charge-neutral WSe$_{2}$ monolayer.}
\newline
Gate-dependent reflection contrast spectra of the bilayer electron system device ($V \mathrm{_{t}} = 4.5$ V). 
The WSe$_{2}$ monolayer has no free carriers as reflected by the absence of charged exciton signal. 
In this situation, $\nu _\text{tot} = \nu _\text{\moire}$ holds, and the incompressible state sensed by the $2s$ corresponds to the Mott insulator (The black arrows coincide in $V \mathrm{_{b}}$).
}
\end{figure*}

\begin{figure*}
\includegraphics[width=1\textwidth]{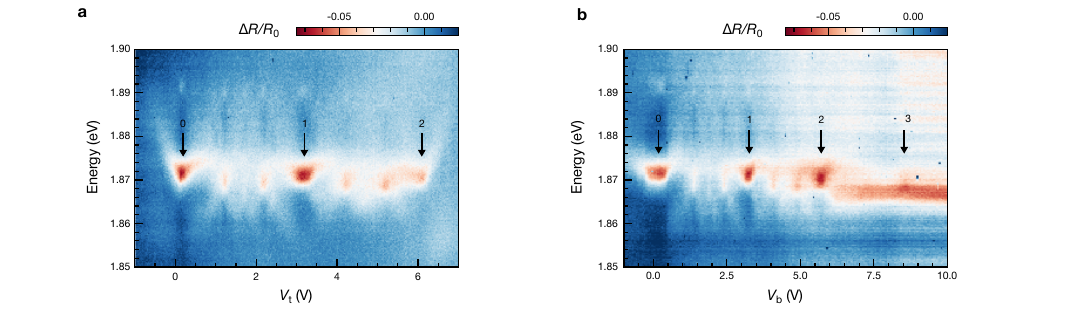}
\caption{\label{figext:zoo}\textbf{Sensing of correlated insulating states at fractional fillings of the moiré superlattice.}
\newline
\textbf{a,} Gate-dependent reflection contrast spectrum with the WSe$_{2}$ monolayer being charge-neutral and acting as sensor layer without taking part in the charge correlations. Starting at $V \mathrm{_{t}} \sim 6$ V, the WSe$_{2}$ monolayer starts charging and the $2s$ exciton disappears. 
\textbf{b,} Gate-dependent ($V \mathrm{_{b}}$) reflection contrast spectrum with the WSe$_{2}$ monolayer being charge-neutral and acting as sensor layer over the whole $V \mathrm{_{b}}$ range. 
The trivial band insulator, the Mott state, the Hubbard band insulator, and the next filled band are labeled by their respective $\nu _\text{\moire}$, that is, 0, 1, 2, 3, respectively.
}
\end{figure*}

\begin{figure*}
\includegraphics[width=1\textwidth]{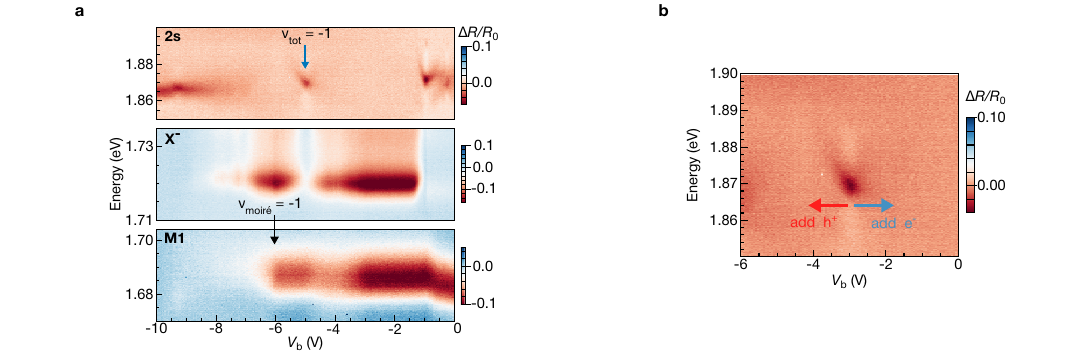}
\caption{\label{figext:hole_side}\textbf{Dipolar insulator and Bose-Fermi mixture around $\nu \mathrm{_{tot}} = -1$.}
\newline
\textbf{a,} Gate-dependent reflection contrast spectrum of the bilayer electron system with both layers being hole-doped ($V \mathrm{_{t}} = -1.2$ V). 
Thedipolar excitonic insulator at $\nu _\text{tot} = -1$ is indicated by the blue arrow, while the Mott insulating state at $\nu _\text{\moire} = -1$ is indicated by the black arrow.
\textbf{b,} Closeup view on the $2s$ resonance from \textbf{a}. The $2s$ exciton exhibits an analog but mirrored asymmetric behavior in $V \mathrm{_{b}}$ as compared to Fig~\ref{fig:asymmetry}b. 
Here, the dipolar excitons are composed of holes in the WSe$_{2}$ monolayer and electrons in the Hubbard band of the moiré heterobilayer.
While adding excess electrons destroys the dipolar excitonic insulating phase, adding holes results in a stable Bose-Fermi mixture. 
}
\end{figure*}

\begin{figure*}
\includegraphics[width=1\textwidth]{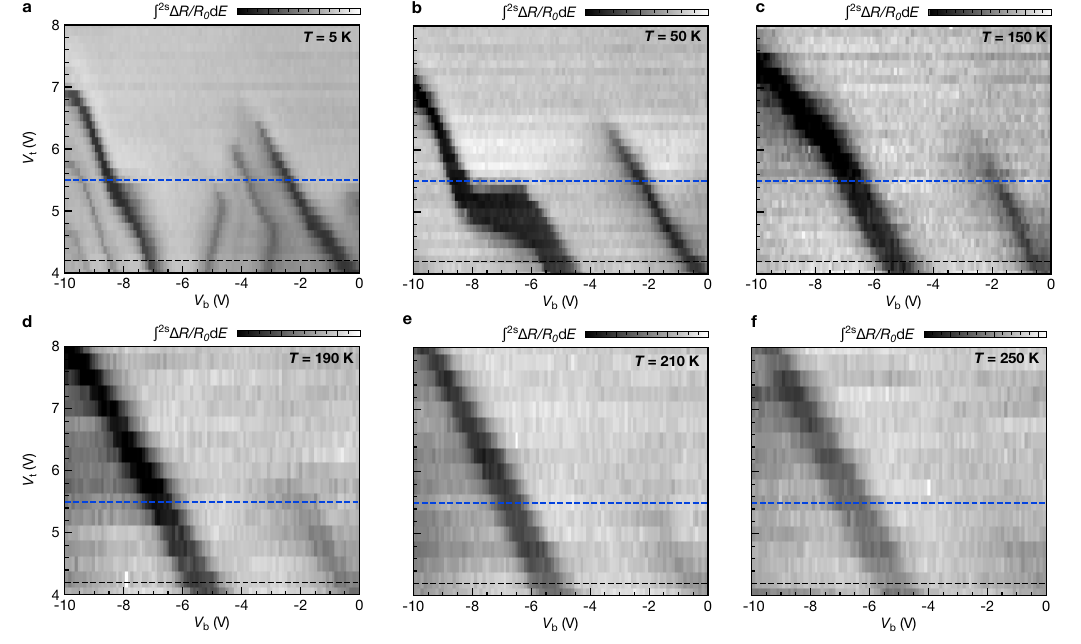}
\caption{\label{figext:temperature}\textbf{Melting temperatures of the excitonic dipolar insulator and the Mott insulator.}
\newline
\textbf{a-f,} Electrostatic phase diagram of the $2s$ exciton at different temperatures in the range 5 - 250 K. We monitor the Mott insulator at $V \mathrm{_{t}} = 4.2$ V (black dashed line) and the dipolar excitonic insulator at $V \mathrm{_{t}} = 5.5$ V (blue dashed line). The Mott insulating state persists up to at least $210$ K, and the dipolar excitonic insulating state shows signatures up to at least $150$ K.
}
\end{figure*}

\begin{figure*}
\includegraphics[width=1\textwidth]{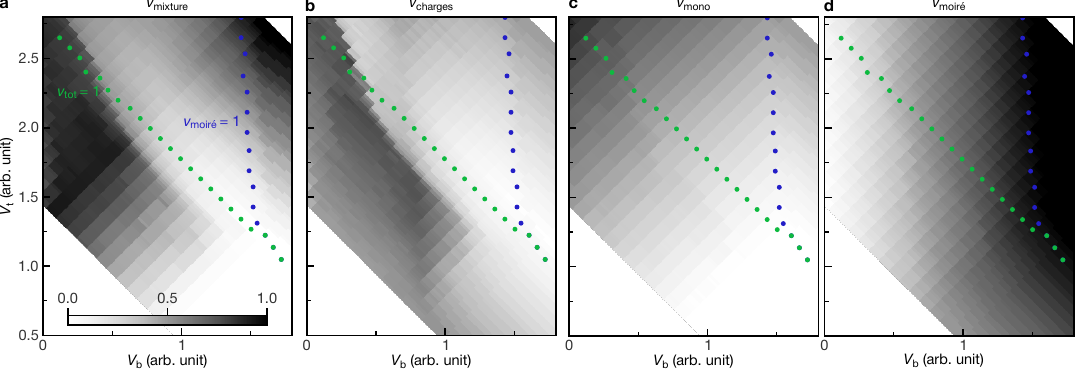}
\caption{\label{figext:theory}\textbf{Mean-field expectation values of densities.}
\newline
Density of unbound fermions in the Bose-Fermi mixture $\nu_\text{mixture}$ (\textbf{a}), total free-charges density from both the mono layer and the \moire layer $\nu_\text{charge}$ (\textbf{b}), electron densities in monolayer $\nu_\text{mono}$ (\textbf{c}) and \moire layer $\nu_\text{\moire}$ (\textbf{d}) as a function of top gate voltage $V_t$ and bottom gate voltage $V_b$. Blue dots indicate a single charge per unit cell $\nu_\text{\moire} = 1$ in the \moire layer, i.e., the Mott insulator, while green dots represent a total filling $\nu_\text{tot}=\nu_\text{\moire}+\nu_\text{mono}=1$.
}
\end{figure*}

\end{document}